\documentclass[12pt]{article}

\newcommand{\ncd}{\newcommand}
\ncd{\dis}[1]{\phantom{\hspace{#1truemm}}}
\ncd{\mc}{\multicolumn}

\newcommand{\plabel}{\label}

\usepackage{epsfig}
\usepackage{amsmath}
\usepackage{amssymb} 
\usepackage{cite}

\newcommand{\beqs}{\begin{equation*}}
\newcommand{\beq}{\begin{equation}}

\newcommand{\eeqs}{\end{equation*}}
\newcommand{\eeq}{\end{equation}}

\newcommand{\beqas}{\begin{eqnarray*}}
\newcommand{\beqa}{\begin{eqnarray}}

\newcommand{\eeqas}{\end{eqnarray*}}
\newcommand{\eeqa}{\end{eqnarray}}

\newcommand{\eq}[2]{\begin{equation} #1 \plabel{#2} \end{equation}}

\newcommand{\eqa}[2]{\begin{eqnarray} #1 \plabel{#2} \end{eqnarray}}

\newcommand{\eps}{\varepsilon}
\newcommand{\al}{\alpha}

\newcommand{\ga}{\gamma}
\newcommand{\de}{\delta}

\newcommand{\la}{\lambda}
\newcommand{\si}{\sigma}

\begin{document}
\begin{titlepage}
\renewcommand{\thefootnote}{\fnsymbol{footnote}}

\hfill TUW--00--02 \\

\begin{center}

\vspace*{\fill}

{\Large\bf The virtual black hole in 2d quantum gravity}\\
  \vspace{7ex}
  D.~Grumiller\footnotemark[1],
  W.\ Kummer\footnotemark[2],
  D.~V.~Vassilevich\footnotemark[3],
  \vspace{7ex}

  {\footnotemark[1]\footnotemark[2]\footnotesize Institut f\"ur
    Theoretische Physik \\ Technische Universit\"at Wien \\ Wiedner
    Hauptstr.  8--10, A-1040 Wien, Austria}
  \vspace{2ex}

  {\footnotemark[3]\footnotesize Institut f\"ur Theoretische 
  Physik \\
Universit\"at Leipzig, Augustusplatz 10, D-04109 Leipzig,\\
Germany}
   \footnotetext[1]{E-mail: \texttt{grumil@hep.itp.tuwien.ac.at}}
   \footnotetext[2]{E-mail: \texttt{wkummer@tph.tuwien.ac.at}}
   \footnotetext[3]{E-mail: \texttt{vassil@itp.uni-leipzig.de}
On leave from Department of Theoretical Physics, St.Petersburg
University, 198904 St.Petersburg, Russia}
\end{center}
\vspace{7ex}
\begin{abstract}
As shown recently (W.\ Kummer, H.\ Liebl, D.V.\ Vassilevich, 
Nucl.\ Phys.\ B {\bf 544}, 403 (1999)) 2d quantum gravity 
theories --- including spherically reduced Einstein-gravity ---  
after an exact path integral of its geometric part can be treated 
perturbatively in the loops of (scalar) matter. Obviously the 
classical mechanism of black hole formation should be contained 
in the tree approximation of the theory. This is shown to be the 
case for the scattering of two scalars through an intermediate 
state which by its effective black hole mass is identified as a 
``virtual black hole''.  The present discussion is restricted to 
minimally coupled scalars without and with  mass. In the first 
case the probability amplitude diverges, except the black 
hole is ``plugged'' by a suitable boundary condition. For massive 
scalars a finite S-matrix element is obtained.
\end{abstract}


\end{titlepage}

\section{Introduction}

Ever since the discovery of Black Hole (BH) evaporation \cite{bhevap} it 
has been evident that quantum processes involving a BH can exhibit
quite unusual properties. In particular, it is not clear
whether such a basic property of the $S$-matrix as unitarity
can be preserved if BHs are present in intermediate
states. At the present level of understanding of quantum
general relativity it seems quite impossible to give a
satisfactory description of such processes in the framework
of full four-dimensional theory. Therefore, the study of
simplified models is so important, among which spherically reduced Einstein
gravity (SRG) is the physically most relevant one. Some
important information can be collected already at the
classical limit. Due to the progress in computer simulations the understanding 
of spherically symmetric collapse of classical matter towards a 
BH has reached a remarkable level \cite{cho93}.  That collapse is 
governed by a critical threshold which even allows a simplified 
discussion in terms of self-similar solutions 
\cite{cho93,fro99}. 
On the other hand, it has already been conjectured \cite{chop} that for (in 
$d=2$) minimally coupled matter (i.e.\ not properly restricted to its s-wave 
part) this critical behaviour is absent, i.e.\ a BH is produced by an 
arbitrarily small amount of matter.

One approach to obtain the quantum version of such processes is the use of 
Dirac quantization for suitably defined operators which describe the collapse 
of matter. A very popular model in this context is the study of thin spherical 
shells. Using the Kucha\v{r} decomposition \cite{kuch92}, especially for null 
(lightlike) shells important progress has been made recently \cite{haj99}. 
We should also mention some earlier papers \cite{early} where the problem
of physical states in dilaton gravity without matter was addressed.
Although certain aspects of the correspondence between physical states and 
BHs could be clarified, the application of these results to the quantum 
$S$ matrix for matter fields was not possible in the framework of the 
reduced phase space formalism.

From the point of view of usual quantum field theory the 
path integral approach seems the most natural one. There 
(properly  defined) S-matrix elements directly determine the 
``physical observables'', which in the Dirac approach are 
--- in the opinion of the present authors --- not so easily 
to be extracted from the ``Dirac observables''. 
During the last years the path integral quantization 
of general 2d gravity theories, including SRG coupled to matter, has shown 
considerable progress \cite{klv97a,klv99}. Based upon the (global and local) 
dynamical equivalence of (torsionless) dilaton theories with first order 2d 
gravity in Cartan variables (with nonvanishing torsion) it turned 
out that the 2d geometry can be integrated much more easily, if a specific 
gauge, the ``light-cone'' gauge for the Cartan variables 
\cite{kum92} is 
chosen. The 2d metric corresponding to this gauge coincides 
with the Eddington-Finkelstein (EF) metric which has the advantage 
of avoiding coordinate singularities at an eventual horizon. As 
shown in \cite{klv99} the explicit one loop contributions in 
the generating functional originate from the Gaussian integral of 
the scalars and from the ``back reaction'' due to the scalars through 
the covariant measure. 

Our present work concentrates on the classical part of this 
functional, i.e.\ on the zero loop order (tree approximation) in 
terms of the scalar matter fields. Clearly here features 
similar to the ones in classical collapse can be expected, 
although those ``macroscopic'' effects could well be modified at 
the ``microscopic'' level, when we refer in the latter case, say, to 
the scattering of individual scalar quanta in 2d gravity theory. 
As will be shown, these effects are indeed present, emerging 
naturally {\em without any further ad hoc assumptions} from the 
2d quantum gravity formalism. 

In order to make this evident we shortly review the main 
arguments of the latter in section 2, generalizing it to the case 
of massive scalars. The classical vertex of scalar fields is 
extracted in section 3. The problems arising for the scattering amplitude 
from our restriction to minimally coupled massless and massive scalars are 
discussed in section 4. In section 5 we summarize our results. 

\section{2d quantum gravity with massive scalars}

\subsection{Path integral quantization}

The total action $L = L^{(g)} + L^{(m)}$ consists of a geometric 
part which is most conveniently written in first order form 
involving Cartan variables 
\begin{eqnarray}
  \plabel{lfirst}
  L^{(g)}&=&\int_{M_2}
\left[ X^+De^-+ X^-De^+ +Xd\omega +\epsilon {\cal{V}}(X^+X^-,X)
  \right] \quad , \\
\plabel{lfirstb}
{\cal{V}}&=&X^+X^-U(X) + V(X) \quad 
\end{eqnarray}
where $De^a=de^a+(\omega \wedge e)^a$ is the torsion two form,
the scalar curvature
$R$ is related to the spin connection $\omega$ by $-\frac{R}{2}=* d\omega$
and $\epsilon$ denotes the volume two form
$\epsilon=\frac{1}{2}\varepsilon_{ab}e^a \wedge e^b=d^2x   \det
e^a_{\mu}=d^2x\,e$.  Our conventions are determined by $\eta=diag(1,-1)$
and $\varepsilon ^{ab}$ by $\varepsilon ^{01}=-\varepsilon ^{10}=1$. We
also have to stress that even with holonomic indices, $\varepsilon^{\mu \nu}$
 is always understood to be the antisymmetric Levi-Civit\'a 
symbol and never the corresponding tensor.  $L^{(g)}$ is globally and 
locally equivalent to the general dilaton theory 
\begin{equation}
 \plabel{lbegin}
L_{(dil)}= \int\, d^2 x \sqrt{-g}
\left(-X\frac{R}{2}-V(X) + \frac{U(X)}2\, (\nabla X)^2\, \right) 
\; ,
\end{equation}
determined by the {\em same} functions $V$ and $U$ of the dilaton 
field $X$ as in (\ref{lfirst}) and (\ref{lfirstb}). In (\ref{lbegin}) 
$g_{\mu\nu}$ is the 2d 
metric, $R$ the Ricci scalar. Spherically reduced gravity (SRG) 
is the special case $U_{SRG} = - (2X)^{-1}, V_{SRG} = - 2$.  

The matter action we write directly in terms of components of the 
zweibeine $e_{\mu}^a$ in $e_{\mu}^ae_{\nu}^b\eta_{ab}$, converting the usual 
expression for the Lagrangian with nonminimally coupled scalar fields $S$ 
(${\cal F}(X) = X$ for SRG)\footnote{A further generalization, 
including selfcouplings is possible without difficulties. Those terms could 
also provide the necessary counterterms for the renormalization when quantum 
corrections to scalar vertices are included. Note, however, that such a 
self-interaction only gives rise to local contributions, while the 
matter-vertices derived by means of our effective theory are non-local in 
general (see below).}
\begin{equation}
\plabel{lmatter}
{\cal{L}}^{(m)} = \frac{{\cal F}(X)}{2} \sqrt{-g}\, \left(g^{\mu \nu}
\partial_{\mu} S \partial_{\nu} S - m^2\, S^2 \right)
\end{equation}
into
\begin{equation}
\plabel{L-matter}
{\cal{L}}^{(m)}  
=- \frac{{\cal F}(X)}{2} \left[ \frac{\varepsilon^{\alpha \mu} 
\varepsilon^{\beta \nu}}{e}
\eta_{ab}e^a_{\mu}e^b_{\nu}\partial_{\alpha}S \partial_{\beta} S 
- m^2 S^2\, e \right] .
\end{equation}
In our paper, as in \cite{klv99}, we treat the simple case ${\cal F} = 1$ of 
minimal coupling. This will be enough to see some of the basic features. 

For the quantum theory --- as well as for the much simplified 
treatment of the exact classical solutions to (\ref{lfirst}) or (\ref{lbegin})
--- the use of the Eddington-Finkelstein (EF) gauge
\begin{equation}
\plabel{gauge-fix}
e_0^+= \omega_0=0  ,\quad      e_0^- =1
\end{equation}
has been found to be useful. It is convenient to introduce the 
shorthand notation for ``coordinates'', ``momenta'' and related sources
\begin{eqnarray}
q_i  &=& (\omega_1, e_1^-, e_1^+) \quad , \nonumber \\
p_i  &=& (X, X^+, X^-) \quad , \plabel{abbrev}\\
j_i  &=& (j, j^+, j^-) \quad , \nonumber \\
J_i  &=& (J, J^-, J^+) \quad . \nonumber
\end{eqnarray}
Following the canonical steps of constructing the path integral 
\cite{klv99}, after integrating out the auxiliary variables and the 
conjugate momentum to $S$ for the gauge (\ref{gauge-fix}) the path integral 
reads 
\begin{equation}
\plabel{W-next}
W=\int\sqrt{\det q_3 }({\cal D}S)({\cal D}^3q)({\cal D}^3p) 
\det F \cdot \exp i\int 
\left( \frac {{\cal{L}}_{(1)}^{\mbox{\scriptsize{eff}}}}{\hbar} +
  {\cal{L}}_{(s)}\right)d^2x 
\end{equation}
where  the  effective Lagrangian, derived from (\ref{lbegin}) and 
(\ref{L-matter}) becomes 
\begin{equation}
\plabel{L-eff-2}
{\cal{L}}_{(1)}^{\mbox{\scriptsize{eff}}} = -q_i\dot{p_i}+q_1 p_2 
-q_3 {\cal V}  - q_2 (\partial_0 S)^2+
(\partial_0 S)(\partial_1 S) - q_3 \frac{m^2}{2}\, S^2\;  .
\end{equation}
It is well known that the correct diffeomorphism invariant
measure for a scalar field $S$ on a curved background $e_\mu^a$
is $d((-g)^{1/4} S) = (d \sqrt{e} S)$, where $e = \det e^a_\mu$ 
\cite{fadxx}. Note, that $ \sqrt{-g} = e = e^+_1=q_3$ in the EF gauge 
is a consequence of the gauge choice (\ref{gauge-fix}).  $\det F$ is the 
determinant resulting from the integration of the auxiliary 
variables of the extended Hamiltonian. It depends on the 
differential operator 
\begin{equation}
\plabel{F}
F = \partial_0 + p_2 U(p_1)
\end{equation} 
The source term includes also the source $Q$ for the scalar 
field as well as sources $J_i$ for the momenta: 
\begin{equation}
\plabel{L-source}
{\cal{L}}_{(s)}= j_i q_i +J_i p_i +SQ \quad .
\end{equation}
Eq. (\ref{L-eff-2}) possesses the crucial property to be linear in the 
``coordinates'' $q_i$. Also the factor $\sqrt{\det q_3}$ in the 
measure may be either lifted into a Lagrangian type term by the 
integration over auxiliary ghost fields that can be integrated 
out later again, or by introduction of an auxiliary metric 
\cite{klv99}. 
In both cases the linearity of the effective action is preserved. 
Thus integrating $d^3 q$ simply produces three $\delta$-functions
\begin{eqnarray}
\plabel{A1}
&{}&\delta\left(-\nabla_0 \left( p_1 -\hat{B_1} \right)\right) \\
\plabel{A2}
&{}&\delta\left(-\nabla_0 \left( p_2 -\hat{B_2} \right)\right) \\
\plabel{A3}
&{}&\delta\left(-F \left( p_3 -\hat{B_3} \right) \right)    \quad .
\end{eqnarray}
where $F$ is defined in (\ref{F}).  $\hat B_i$ are functions of the 
sources $j_i$ and matter fields and will be given below. Using 
these three $\delta$-functions the integrations over $(d^3p)$ 
yield directly 
\begin{equation}
\plabel{pi}
p_i=\hat{B}_i \quad .
\end{equation}
This simply means that in the phase-space (path-) integral
only classical paths contribute to the $p_i$ and the remaining continuous
physical degrees of freedom are represented by the scalar field alone, since
all integrations over geometric variables have been performed exactly.  
Note that integration over $p_3$ from (\ref{A3}) produces another factor 
$(\det F)^{-1}$ so that the total Faddeev-Popov determinant is one --- 
a result consistent with experience from Yang-Mills fields in 
temporal gauges like (\ref{gauge-fix}). 

By $\nabla_0 = \partial_0 + i\, (\mu + i \varepsilon )$ we define a 
regularized time derivative  with $\epsilon, \mu \to + 0$ for 
describing the IR and UV regularized one-dimensional associated 
Green function in loop integrals. Homogeneous 
modes always appear when we invert the operator 
$\nabla_0$. Such modes $( \nabla_0\, \bar p_i = 0)$ must be 
included in $\hat B_i$ where they completely describe the 
(eventual) classical background \cite{klv99}\footnote{E.g. for SRG that
background may be a BH or flat Minkowski spacetime, depending on the choice of
modes $\bar{p}_i$. The freedom still 
encoded in the homogeneous modes can be reduced by fixing the 
residual gauge freedom of the EF gauge (6) and solving Ward 
identities. This procedure is described in detail in 
\cite{klv99}. }.
\begin{eqnarray}
\hat{B}_1 &=& \underbrace{\bar{p}_1 + \nabla_0^{-1} \bar{p}_2 + 
\hbar (\nabla_0^{-1}j_1 + \nabla_0^{-2}j_2) } 
 -\nabla^{-2}_0 (\partial_0 S)^2 \; ,
\plabel{B12} \\
\nonumber {}&{}& \qquad \qquad \qquad:=B_1 \\
\hat{B}_2 &=& \underbrace{\bar{p}_2 + \hbar \nabla_0^{-1} j_2 }  
-\nabla^{-1}_0 (\partial_0 S)^2 \; ,\quad \plabel{B22} \\
\nonumber {}&{}& \qquad :=B_2
\end{eqnarray}
\begin{eqnarray}
\plabel{B32}
\hat{B}_3 &=& e^{-\hat{T}} 
\left[ \nabla_0^{-1}e^{\hat{T}}(\hbar j_3 -V(\hat{B}_1)
-\frac {m^2}2 S^2)+\bar{p}_3
\right] \\
{}        &=& \underbrace{
e^{-T}\left[ \nabla_0^{-1}e^{T}(\hbar j_3 -V(B_1))+\bar{p}_3 \right]}   
+{\mbox{terms }}{\cal{O}}(S^2) \; . \\
{}        &{}& \qquad \qquad \qquad :=B_3 \nonumber
\end{eqnarray}
\begin{equation}
\hat{F} = e^{-\hat{T}}\nabla_0 e^{\hat{T}}, \quad  
\hat{T} = \nabla^{-1}_0 (\hat{U} \hat{B}_2), \quad 
\hat{U} = U(\hat{B}_1) \plabel{Uhat} \quad .
\end{equation}
In the abbreviations an exponential representation for the 
operator $F$ is used. 
There is still an ambiguity in the path integral. Indeed, the 
term $\int J_3 \hat B_3$ can be formally rewritten as
\begin{equation}
\int\; \left( e^{\bar T}\, (\hbar j_3 - V ( \hat B_1) - 
\frac{m^2}{2} S^2 ) ( - \nabla_0 ) e^{-\hat T}\, J_3 + J_3 \bar 
p_3\; \right) \plabel{21}
\end{equation}
We have a freedom to add a homogeneous solution $\nabla_0 \tilde 
g = 0$ to the term $e^{-\hat T}\, J_3$. This amounts to adding to 
the effective Lagrangian the term 
\begin{equation}
\plabel{L-HK}
\tilde{\cal{L}}=\tilde{g} e^{\hat{T}} \left(\hbar j_3 - \hat{V} - 
\frac{m^2}{2}\, S^2 \right)\, . 
\end{equation}
The same procedure applied to the terms $J_1 \hat B_1$ and $J_2 
\hat B_2$ just leads to trivial contributions. Clearly $\tilde{\cal L}$
alone survives when the sources $J_i$ for the 
momenta are switched off. Nevertheless, those sources are 
technically important for a simple definition of an overall conservation 
law $d {\cal C} = 0$, peculiar to all $2d$ theories, even with 
interacting matter \cite{kum95}. Its geometric part $ ( Q = \int^{p_1} U 
(y) dy)$ 
\eq{
{\cal C}^{(g)} = e^{Q(p_1)} p_2 p_3 + \int^{p_1} V(u) e^{Q(u)}du
}{23}
for SRG by fixing integration constants in a specific way may be defined as
\eq{
{\cal C}^{(g)}_{SRG} = \frac{p_2p_3}{\sqrt{p_1}} - 4 \sqrt{p_1} 
}{24}

\subsection{Effective scalar theory}

Having performed the integral $({\cal D}^3 q)$ and then $({\cal D}^3 p)$ which 
is possible only in the chosen gauge (\ref{gauge-fix}) in a straightforward 
way, the generating functional (\ref{W-next}) becomes \cite{klv99}
\eq{
W(j,J,Q) = \int ({\cal D}S)\exp{i \int d^2x \left( \frac{{\cal L}^{eff}_{(2)}}
{\hbar} + SQ \right)} .
}{25}
Scalar fields can be integrated only perturbatively. Let us 
separate different orders of $S$ in the effective Lagrangian:
\begin{equation}
{\cal L}^{eff}_{(2)} = {\cal L}_0 + S Q + {\cal L}_2 + {\cal 
L}^{int}
\plabel{26}
\end{equation}
where ${\cal L}_0$ does not contain $S$ and ${\cal L}_2$ is quadratic in 
$S$. All higher powers are collected in ${\cal L}^{int}$. 
According to (\ref{25}) the quadratic part ${\cal L}_2$ describes a 
free minimally coupled scalar field on the effective background 
geometry with the zweibein expressed in terms of the external 
sources ($T = \hat{T}(S = 0)$):
\eq{
{\cal L}_2 = \left(\partial_0 S\right)\left(\partial_1 S\right) - E_1^-\left(
\partial_0 S \right)^2 - \frac{m^2}{2}S^2E_1^+
}{27}
\eq{
E_1^+ = e^T, \hspace{0.5cm} E_1^- = - \partial_0^{-2} \tilde{g} e^T \left(
V' + UV \right)
}{28}
We use capital letters $E_1^\pm (J, j)$ to distinguish the effective 
values from the fundamental zweibein fields $e^\pm_1$ which  are already 
integrated out.

The interaction Lagrangian can be represented as
\begin{equation}
{\cal L}^{int} (S) \to {\cal L}^{int}\; \left( 
\frac{1}{i}\frac{\delta}{\delta Q} \right)
\plabel{29} 
\end{equation}
and pulled out from the integral over $S$. As shown in \cite{klv99} the path 
integral measure for $S$ by a straightforward redefinition can be 
reduced to just the standard Gaussian one. In the generating functional 
\eqa{
W &=& \exp{\left\{\frac{i}{\hbar}\int d^2x{\cal L}^{int} \right\}} \nonumber \\
&& \times \exp\, \left\{ \frac{i \hbar}{2} \; \int_x \int_y Q(x) G_{xy}\, 
Q (y) + \int_x (J_iB_i) + i \Gamma^{1-loop} (j, J) \right\} 
}{30}
where $G_{xy}$ is the scalar field propagator on the effective 
background (\ref{28}) and $\Gamma^{1-loop}$ is the logarithm of the 
determinant which for $m = 0$ may be expressed as a Polyakov 
action. We do not have to go into details on the one-loop 
contributions since the present paper deals with the tree-level 
diagrams which are determined by the first two terms in (\ref{30}), where 
${\cal L}^{int}$ is to be interpreted as in (\ref{29}). 

\section{Vertices of scalar field}

In our present paper of primary interest are the effective scalar vertices 
contained in ${\cal L}^{int}$. At vanishing sources $J_i = 0$ for the 
momenta, ${\cal L}^{int}$ in (\ref{26}) reduces to the corresponding 
contribution from $\widetilde {\cal L} $ as defined in (\ref{L-HK}). 

\subsection{Massless scalars}

Let us first consider the case $m = 0$. Then the scalar field $S$ enters 
the interaction Lagrangian only as $(\partial_0 S)^2$. Moreover, 
according to (\ref{pi}) and (\ref{B12}), which, in turn, are the input in 
(\ref{B32}) and (\ref{L-HK}), it always appears in the combination 
$\lbrack \hbar j_2 - (\partial_0 S)^2 \rbrack $ . Therefore, the effective 
vertex of  order $2n$ in ${\cal L}^{int}$ of (\ref{26}) has the generic 
form\footnote{From now on we put $\hbar = 1$ for simplicity.}: 
\begin{equation}
S^{(2n)} = \int d^2 x_1 \ldots \; d^2 x_n\, S^{(2n)} (x_1, \ldots 
, \, x_n ) (\partial_0 S)^2_{x_1} \ldots (\partial_0 S)^2_{x_n}
\plabel{31}
\end{equation}
where
\begin{equation}
S^{(2n)}= \left. \frac{(-1)^n}{n!}\; 
\frac {\delta^n}{\delta j_2^n} \, \tilde{\cal L}\right\vert_{j=0}
=\left. \frac {(-1)^{n-1}}{(n)!}\, \left( \frac {\delta^{n-1}}
{\delta j_2^{n-1}}\, E^-_1 \right)\, \right\vert_{j=0}
\plabel{b88}
\end{equation} 
with $E_1^-$ defined in (\ref{28}). 

To obtain the $(n-1)^{th}$ functional derivative of $E_1^-$ it is enough to
take $j_2$ localized at $n-1$ different points:
\begin{equation}
j_2(x) = \sum_{k=1}^{n-1} c_k \delta (y_k-x)
\plabel{f88}
\end{equation}
then we can expand $E_1^-(j_2,x)$ in a power series of $c_k$.
In the resulting sum the coefficient of the term with 
$\prod_{k=1}^{n-1}\, c_k$ will give the desired
functional derivative. 

As  seen from (\ref{Uhat}), (\ref{28}) $E^-_1$ is a nonlocal functional
of the $p_i$ which are again nonlocal functionals of the sources 
$j_i$. Our aim is the determination of the classical vertices. 
Therefore, their regularizations introduced in $\nabla_0$ may be 
removed and $\partial_0^{-1}$ simply becomes an integration $\int 
dx_0$. However, 
instead of applying this integration  several
times it is more convenient to solve the corresponding differential
equations with suitable boundary conditions. This may seem 
surprising, because (\ref{b88}) in principle already represents the 
solution in closed form. But the treatment of multiple 
integrations is very involved with many, at first, undetermined integration 
ranges and integration constants, which --- as for the BH in SRG --- have 
singularities. Also the trick to go back to the 
classical equations which determine $\widetilde{\cal L}$ (cf.\ 
\cite{klv99}) will give us important additional information to be 
used for the physical interpretation of the results. 

Our starting point are the three differential equations for $p_i$ which follow
from solving the $\de$-functions (\ref{A1})-(\ref{A3}) 
(cf.\ eqs.\ (\ref{39}-\ref{41}) of \cite{klv99}) in the presence of a source 
term  $j_2 q_2$ in $\cal L$ whose solution is (\ref{pi}), in the special 
case $S = 0, j_1 = j_3 = 0$, i.e.\ $(\partial_0 a = \dot a)$: 
\begin{eqnarray}
\dot p_1 - p_2 & = & 0 \plabel{34} \\
\dot p_2 & = & j_2 \plabel{35} \\
\dot p_3 + p_2 U p_3 + V & = & 0 \plabel{36}
\end{eqnarray}
with $j_2$ of (\ref{f88}). The quantity $E_1^- = q_2 (j_2)$ in the 
notation (\ref{abbrev}) may be calculated from the classical e.o.m.'s for 
$q_i$ \cite{klv99} or, equivalently, by suitable differentiations of 
(\ref{28}) ($U^\prime = dU/dp_1$ etc.\ ): 
\begin{eqnarray}
\dot q_1 - p_2p_3q_3 U^\prime - q_3 V^\prime & = & 0 \plabel{37} \\
\dot q_2 + q_1 - q_3 p_3 U & = & 0 \plabel{38} \\
\dot q_3 - p_2 q_3 U & = & 0 \plabel{39}
\end{eqnarray}
From (\ref{38}) and (\ref{39}), eliminating $\dot q_1$ by (\ref{37}), or 
directly differentiating twice (\ref{28}) the simple differential equations 
for $q_2 = E_1^-$ may be obtained: 
\begin{equation}
{\ddot q}_2 + q_3 \, (V^\prime + U V ) = 0
\plabel{40}
\end{equation}
where $q_3 = E_1^+$ is already determined by the first Eq. (\ref{28}). 

\begin{figure}
\centering
\epsfig{file=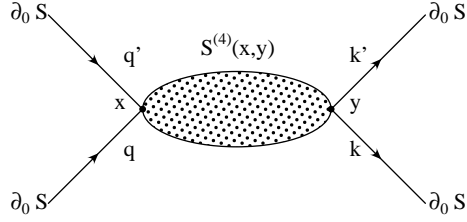,height=3cm}
\caption{$S^{(4)}$-vertex with outer legs}
\plabel{fig1}
\end{figure}

In the following we restrict ourself to the vertex $S^{(4)} (n = 
2)$ in (\ref{31}). It is depicted in Fig. 1 (the momenta and corresponding 
arrows we included for later reference in the basic $S$-matrix elements with
Minkowski modes). Then in (\ref{f88}) only one term with $c_k = c$ is needed 
and 
\eq{
S^{(4)}(x,y) = - \frac{1}{2}\frac{\de q_2(x)}{\de j_2 (y)} .
}{41} 

In $S^{(4)}$ only the part of (\ref{41}) contributes that is symmetric in $x 
\leftrightarrow y $. For $E_1^-$ in (\ref{40}) the only input 
from the momentum equation are the contributions to $p_1$ and 
$p_2$. Fixing the residual gauge transformations these solutions 
with (\ref{f88}) at $k =1$ can be written as $(x^0 = t, y^0 = s)$
\eq{
p_2 = 1 + c \left[ \al + \Theta(t-s)\right]
}{42}
\eq{
p_1 = t + c \left[ \al + \Theta(t-s)\right](t-s)
}{43}
where one integration constant has been absorbed in the definition of $t$.
An overall factor  $\delta (x^1 - y^1) $ of the square brackets 
in (\ref{42}) and (\ref{43}), expressing the locality in $x^1$, is not 
written explicitly but will be taken into account in the end. The 
constant\footnote{Here and in the following a `constant' means 
also functions of $x^1$ and $s=y^0$.} $\alpha$ parametrizes different 
possible solutions. There are three main cases: \\ 
\begin{tabular}{rll}
\mc{1}{l}{\dis{15}} & \mc{1}{l}{\dis{50}}& \mc{1}{l}{\dis{30}} \\
a) &  $p_2 \neq 1 $ for $t > s$ only: & $\alpha = 0$\\
b) &  $p_2 \neq 1 $ for $ t < s$ only: & $\alpha = -1$ \\
c) & ``symmetric'' solution: & $\alpha = - \frac12 $ .
\end{tabular} \\

In the last case the square bracket in (\ref{42}) and (\ref{43}) may be 
replaced by half the sign function $\varepsilon (t -s)$. Although 
$p_3$ does not enter eq. (\ref{40}), it is necessary to compute the effective 
BH mass. From the general solution of (\ref{36})
\eq{
p_3 = e^{-Q} \left( \bar{p}_3 - \int^{p_1} du V(u) e^{Q(u)} \right)
}{44}
with  $\dot {\bar p}_3 = 0$ for each of the cases a) b) c) it is 
simple to find a solution with $p_3^{t \leq s}(t=s) = p_3^{t \geq s}(t=s)$.
The most interesting application is SRG with $U_{SRG} = - (2p_1)^{-1}, 
V_{SRG} = - 2$. E.g.\ for case a) one finds easily with (\ref{34}) for 
${\cal C}^{(g)}_{SRG}$
\eq{
\left.{\cal C}^{(g)}\right|_{t < s} = \bar{p}_3, \hspace{0.5cm}
\left.{\cal C}^{(g)}\right|_{t > s} = \bar{p}_3 + c \left( \bar{p}_3 + 
4 \sqrt{s}\right)
}{45}
Here the integration constant $\bar p_3$ must be independent of $c$. Thus in 
the term $O (c)$, relevant for our vertex $S^{(4)}$ a 
nonvanishing effective ``BH mass'' has to be present. Since 
${\cal C}^{(g)} \propto - m_{BH}$ (cf. e.g.\ the last ref.\ \cite{kum95}) 
with a ``natural'' choice $\bar p_3 = 0$ for the solution without source 
$j_2$, a BH mass proportional $(- \sqrt{4}{s})$ will be switched 
on for $t > s$. As will be clarified below --- despite our 
suggestive notation ---  $t$ and $s$ refer to a space coordinate. 
Our gauge choice (\ref{42}) and (\ref{43}) for SRG has placed the singularity 
at $t =0$ which, however, in this case would not lie in the region $t > s$ 
where ${\cal C}^{(g)}$ differs from zero. Thus case a) suggests the 
interpretation of a shell with negative mass, situated at $t > s$. 

For case b) an analogous computation gives $m_{ADM} \propto + 4\sqrt{s}$ for 
$t < s$ only, i.e.\ a proper BH at $t = 0$, whose effect is switched off for 
$t \ge s$. For c) ${\cal C}^{(g)}$ jumps from $- 2\sqrt{s}$ at $t < s$ to 
$+ 2\sqrt{s}$ at $t > s$. The common feature of this apparently highly 
ambiguous situation (also other values of $\alpha$ may be taken in (\ref{42}) 
and (\ref{43})!) is the discontinuity in the effective BH mass at $t = s$ 
which will make the appearance of the a singularity in $S^{(4)}$ 
unavoidable. We call this phenomenon ``virtual Black Hole'' (VBH). 
Actually the ambiguity in $\alpha$ {\em disappears alltogether} 
in the vertex $S^{(4)} (x,y) = S^{(4)} (y,x)$ so that the 
different interpretations in a), b), c) and for other values for 
$\alpha$ should not be taken at face value. It should be noted 
that the range of variables $t = x^0, s = y^0$ are not to be 
identified as the variables to be used in a scattering amplitude 
$S + S \to S + S$ connecting asymptotic Minkowski space scalar 
fields (see below). 

We now turn to the solution of (\ref{39}) and (\ref{40}), using case a) for 
(\ref{42}) and (\ref{43}) in anticipation of the fact that it will be 
symmetrized anyhow to the only relevant contribution to 
$S^{(4)}$. On the other hand, for another vertex to appear below 
for massive scalars, only that case will produce a finite result.

The continuous solution to (\ref{39}) for SRG is (the indices 
$\scriptstyle (0)$ and $\scriptstyle (1)$ refer to $t < s$ and $t 
> s$, respectively; the integration constant in $Q$ is fixed 
according to $Q_{SRG} = T_{SRG} = -\frac12 \ln t) $
\eqa{
q_3^{(0)} &=& \frac{\bar{q}_3}{\sqrt{t}}, \plabel{46} \\
q_3^{(1)} &=& \frac{\bar{q}_3}{\sqrt{t}} \left[ 1 - c \left(1-\frac{s}{t}
\right) \right] .
}{47}
Introducing (\ref{46}), (\ref{47}) into (\ref{40}) for continuous $q_2$ and 
$\dot q_2$ at $s = t$
\eqa{
q_2^{(0)} &=& \bar{q}_3 \left[ 4 \sqrt{t} - \frac{2t}{\sqrt{s}} - 2\sqrt{s} +
\bar{a}t + \bar{b} \right] , \plabel{48} \\
q_2^{(1)} &=& \bar{q}_3 \left[ \frac{4}{(1+c)^2} \sqrt{t(1+c)-sc} + \frac{2}
{1+c} \left( \sqrt{s} - \frac{t}{\sqrt{s}}\right) \right. \nonumber \\
&&\left. - \frac{4\sqrt{s}}{(1+c)^2} + \bar{a}t + \bar{b} \right] , 
}{49}
there is still a dependence on integration constants $\bar q_3, 
\bar a, \bar b$. It should be noted that $\bar q_3$ in a direct 
calculation from $\tilde{\cal L}$ in (\ref{21}) would be replaced by 
the factor $\tilde g (x^1)$. 

Next we will see that these constants can be fixed 
uniquely by natural assumptions for the effective line element, 
computed in the gauge (\ref{gauge-fix}) from $q_2 = E_1^{-}, q_3 = E_1^{+}$ 
(we set $x^1 = x$): 
\begin{equation}
(ds)^2 = 2 q_3 (dt + q_2 dx ) dx
\plabel{50}
\end{equation}
For $t < s$ with (\ref{46}) and (\ref{48}) in case a) we require the line 
element to describe flat (Minkowski) space, i.e.\ with a new coordinate 
$\bar t$
\eq{
(ds)^2_{(0)} = 2 d\bar{t}dx+(dx)^2 = (d\tau)^2 - (dz)^2
}{51}
This completely (only up to a sign in $\bar q_3$ which we 
chose to be positive) fixes 
\eqa{
b = sa = 2\sqrt{s}, \vspace{0.5cm} \bar{q}_3 = \frac{1}{2\sqrt{2}}, \plabel{52}
\\
\sqrt{2}\bar{t} = \sqrt{t}
}{53}
Otherwise a BH and an acceleration term (Rindler 
metric) would be present. In the last equality (\ref{51}) the 
transition from (outgoing) EF coordinates to usual Minkowski coordinates 
\begin{equation}
x = \tau - z, \qquad \bar t = z \plabel{54}
\end{equation}
has been made. The relations (\ref{52}) also lead to a unique result 
for the term of first order in $c$ which determines 
\begin{equation}
\frac{\delta q_2 (x)}{\delta j_2 (y)} = - \frac{1}{2 \sqrt{2}}\; 
\frac{\vert \sqrt{x^0} - \sqrt{y^0} \vert^3}{\sqrt{x^0 y^0}} \de (x^1 - y^1 ) .
\plabel{55}
\end{equation}
Here the symmetrization has been performed and the overall factor 
$\delta (x^1 - y^1)$ included. This result is proportional to the
(symmetrized) vertex calculated in \cite{klv99}, where the overall constant
had not been determined.

It is also essential to study the effective line element $(ds)^2_{(1)}$ valid
in the range $t \ge s$. Again we may first bring it into EF form in terms of 
a new coordinate $\bar{t}$. Joining $\bar{t}$ smoothly to the corresponding
variable for $t \le s$ we get
\eq{
\bar{t}_{(1)} = \frac{1}{\sqrt{2}} \left[ \sqrt{t} - \frac{1}{\sqrt{t}} + 
\sqrt{s}+c(\sqrt{s}-\sqrt{t})\right]
}{55n}
which for large $t$ and $\bar{t}_{(1)}$ reduces to $\bar{t}_{(1)} \to \left(
\sqrt{t}(1-c)+\sqrt{s}(1+c)\right)/\sqrt{2}$. In the line element
\eq{
(ds)^2_{(1)} = 2d\bar{t}_{(1)}dx+K_{(1)}(dx)^2
}{56n}
the Killing norm $K_{(1)}$ in terms of $\bar{t}_{(1)}$ (for case a)) can be
calculated easily to the required ${\cal O}(c)$ from
$K_{(1)} = 2q_2^{(1)}q_3^{(1)}$ for asymptotic values of the radial variable
$\bar{t}_{(1)}$:
\eq{
\lim_{\bar{t}_{(1)} \to \infty} K = 1 + c \left[ \frac{3\sqrt{s}}{2\sqrt{2}
\bar{t}_{(1)}} + \frac{\bar{t}_{(1)}}{\sqrt{2s}} - 3 + {\cal O}\left(
\frac{1}{\bar{t}^2_{(1)}}\right) \right]
}{57n}
The first term in the square bracket, in agreement with (\ref{45}), describes
the VBH as a massive object with (negative) effective mass proportional to
$\sqrt{s}$. Eq. (\ref{57n}) is valid asymptotically, but the linear dependence
on $\bar{t}_{(1)}$ indicates that it corresponds to a uniformly accelerated
coordinate system with respect to Minkowski space. The acceleration is 
proportional to $s^{-1/2} = (y_0)^{-1/2}$. Thus in the simultaneous limit
with $y_0 \to \infty$ the asymptotic scalar fields, entering an $S$-matrix 
element to be computed from (\ref{55}) in a certain sense may be determined by 
Minkowski modes after all.

It is obvious that similar arguments for case b) will produce flat Minkowski 
space for $t > s$. At $t < s$ something like a 
genuine BH again together with linear terms in the radial variable 
$\bar{t}_{(1)}$ appears. However, the restriction of the BH-like structure to 
the interval $0 \le t \le s$, at least for any finite $s$ does not permit the 
definition of an asymptotic radial variable associated with some corresponding 
Rindler space. The ``symmetric'' case c) and all other situations with general 
$\al$ in (\ref{42})-(\ref{43}) have Rindler terms in the whole range of $t$. 
Thus the presence of an asymptotically flat Minkowski space on the sense of a
double limit $x^0 \to \infty, y^0 \to \infty$ (subjected to a very special
sequence of those limits) is restricted to case a). 

\subsection{Massive scalars}

\begin{figure}
\centering
\epsfig{file=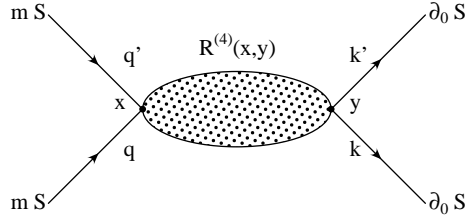,height=3cm}
\caption{$R^{(4)}$-vertex with outer legs}
\plabel{fig2}
\end{figure}

For massive scalars the last term in (\ref{L-HK}) again occurs in 
combination with a source for the zweibeine, namely $j_3$. For 
the vertex $S^{(4)}$ now another term is created from an 
$(\partial_0 S)^2$ in $e^{\hat T}$ and that mass term. Therefore in 
the new vertex contribution
\eq{
R^{(4)} = \int d^2x \int d^2y S_x^2 R^{(4)}(x,y) \left( \partial_0 S
\right)^2_y
}{56}
we get $(\hbar = 1)$
\eq{
R^{(4)} = - \frac{m^2}{2} \frac{\delta^2 \tilde{{\cal L}}}{\de j_3(x) 
\de j_2 (y)} = -\frac{m^2}{2} \frac{\de q_3(x)}{\de j_2(y)}
}{57}

This expression is simply the factor of $c$ in (\ref{47}). Together 
with the normalization of $\bar q_3$ and the overall $\delta (x^1 
- y^1)$ we obtain in case a) of our different solutions for (\ref{42}) 
and (\ref{43})
\eq{
\frac{\de q_3(x)}{\de j_2(y)} = - \frac{\Theta(x^0-y^0)}{2\sqrt{2}(x_0)^{3/2}}
(x_0-y_0) .
}{58}
The crucial difference to the other cases b) and c) consists in the 
property of (\ref{58}) that only here the step function ``protects'' the 
$x$ and $y$ integrations from the singularity at $x^0 = 0$ in $R^{(4)}$ (see 
below).

\section{Scattering Amplitude}

For the scattering process of two scalars $S + S \to S + S$ 
through the two vertex contributions $S^{(4)} + R^{(4)}$ of the 
previous section we first transform both $x$ and $y$ to the asymptotically 
flat coordinates (\ref{54}). Clearly, in view of the remarks after (\ref{57n}),
it may seem questionable that this is consistent with the properties of the
vertex at asymptotic distances. Only in case a) an effective line element, say
in $x$, exists which is asymptotically flat. But it refers to an accelerated
system whose acceleration is proportional to $(y_0)^{-1/2}$. On the other hand,
the free fields $S(x)$ in the interaction picture approach to standard
scattering theory cannot show a dependence on the variable $y$ of a different
vertex. This, in our opinion, justifies the assumption that (for case a)) there
must be an asymptotic limit towards ``independent'' flat Minkowski space for
$S(x)$ and $S(y)$ for a (hopefully) gauge fixing independent $S$-matrix 
element to exist. Then (\ref{31}) with (\ref{41}) and (\ref{55}) 
yields $(x^1 = \bar x^1)$ the manifestly nonlocal vertex
\eq{
S^{(4)} = \frac{1}{64} \int d^2\bar{x} \int d^2\bar{y} \Theta(\bar{x}_0)
\Theta(\bar{y}_0) \frac{\left( \partial_{\bar{x}_0} S\right)^2}{\bar{x}_0^2}
\frac{\left( \partial_{\bar{y}_0} S\right)^2}{\bar{y}_0^2} 
\left| \bar{x}_0 - \bar{y}_0 \right|^3 \de(x_1 - y_1)
}{60}
which is the same for $m = 0$ and $m \neq 0$ as well as for all possible
$\al$-prescriptions.

In a similar manner the additional vertex for massive scalars 
from (\ref{gauge-fix}), (\ref{57}) and (\ref{58}) becomes
\eqa{
R^{(4)} &=& -\frac{m^2}{8} \int d^2\bar{x} \int d^2\bar{y} \Theta(\bar{x}_0)
\Theta(\bar{y}_0) \de(x_1-y_1) \nonumber \\
&& \vspace{2cm} \times \frac{\Theta(\bar{x}_0-\bar{y}_0) 
(\bar{x}_0^2-\bar{y}_0^2)}{\bar{x}_0^2\bar{y}_0} S^2_{\bar{x}_0}
\left( \partial_{\bar{y}_0} S\right)^2 .
}{61}
The vertex has been given for the case a) ($\alpha = 0$) in (\ref{43}).

\subsection{Massless Scalars}

Evidently both vertices (\ref{60}) and (\ref{61}) exhibit singularities at 
$\bar x^0 = 0, \bar y^0 = 0$. Let us consider asymptotic massless 
scalars first. Without imposing any boundary condition at ${\bar 
x}^0 = 0$ the usual decomposition of the scalar field into Minkowski space
modes will be 
\eq{
S = \frac{1}{\sqrt{2 \pi}} \int \frac{dk}{\sqrt{2k}} \left(a^+_R e^{ik(\tau-z)}
+ a^+_L e^{ik(\tau+z)} + a^-_R e^{-ik(\tau-z)} + a^-_L e^{-ik(\tau+z)}\right) ,
}{62}
where the indices $R$ and $L$ denote the right and left moving parts 
in our one-dimensional situation. In the ``outgoing EF 
coordinates'' (\ref{54}) used in (\ref{gauge-fix}) the arguments are simply 
$\tau - z = \bar{x}^1, \tau + z = \bar{x}^1 + 2\bar{x}^0$ . The singularity of 
(\ref{63}) in a scattering  matrix element for two ingoing and outgoing 
Minkowski quanta of the $S$-field with momenta 
$q, q'$ and $k, k'$, respectively
\eq{
T(q, q'; k, k') = \frac{1}{2} \left< 0 \left| a^-(k)a^-(k')S^{(4)}a^+(q)a^+(q')
\right| 0 \right>
}{63}
for any $R$ and $L$ can be interpreted as an indication that in 
this case the formation of a BH is ``inevitable''. Indeed 
regularizing (\ref{60}) with $({\bar x }^0)^2 \to \lim_{\delta \to 0} 
({\bar x }^0\,{}^2 + \delta^2)^{-1} $ and defining a left 
moving wave packet formally by $S \to \delta^{3/4}\, S$ would 
yield  a finite result (up to an undetermined factor). The observed divergence 
of the scattering amplitude (\ref{63}) for our vertex which contributes 
to the tree approximation thus seems to be in qualitative agreement 
with the conjecture \cite{chop}, when scalar fields with minimal 
coupling are introduced at the reduced ($1+1$) level: Then the BH is formed 
for arbitrary small amounts of collapsing matter. A threshold for BH 
formation only occurs if nonminimally coupled scalars are 
considered, corresponding to proper taking into account of the 
$s$-wave nature in the spherically reduced situation \cite{cho93}.

A radical solution to the divergence problem consists in imposing a suitable 
boundary condition\footnote{This boundary condition
implies that $S$ becomes a self-dual scalar field at the origin - an
essential difference to Dirichlet or Neumann boundary conditions, which - 
in a certain sense - are dual to each other \cite{limi98}.} for the scalar 
field to make (\ref{gauge-fix}) finite:
\begin{equation}
\left. \frac{\partial S}{\partial {\bar x}^0}\; 
\right\vert_{{\bar x}^0 = 0} \; = \; \left( \frac{\partial 
S}{\partial \tau} + \frac{\partial S}{\partial z} \right) = 0\; .
\plabel{64}
\end{equation}
However, this eliminates the left-movers $( a_L^\pm = 0)$ in (\ref{60}) 
also at ${\bar x}^0 \neq 0$, and (\ref{61}), as well as in all higher 
order vertices. The physical system now 
consists of a background, eventually describing a fixed BH by a singularity at
${\bar x}^0 = z = 0$, and free right-moving scalars which run away 
from it. No genuine BH formation occurs in this setting. 

\subsection{Massive scalars}

Although also in this case half of the modes are eliminated, a nontrivial 
result is obtained for free massive scalars with 
energy $E_k = \sqrt{k^2 + m^2}$, obeying the boundary condition 
(\ref{64}). The modes can be extracted from
\begin{eqnarray}
S & = & \int\limits_0^\infty dk\; N (k)\, \left\{ a^+ (k) \, e^{i E 
({\bar x}^0 + {\bar x}^1)} \lbrack (E_k + k) \, e^{- ik {\bar x} 
^0} - \right. \\ \nonumber
&& \qquad \left. - (E_k - k) \, e^{ik {\bar x}^0 }\rbrack + h.c. 
\right\} \plabel{65}
\end{eqnarray}
with the normalization factor 
\eq{
N(k) = \left[4 \pi E_k (E_k^2+k^2)\right]^{-1/2}
}{66}
determined such that the Hamiltonian is $ H = \int_0^\infty dk E_k 
a^+ a^-$.

Now
\eq{
\frac{\partial S}{\partial \bar{x}_0} = 2m^2 \int_0^{\infty} dk N(k) 
\sin{k\bar{x}_0} \left( a^+e^{iE_k(\bar{x}_0+\bar{x}_1)} + h.c. \right)
}{67}
clearly obeys (\ref{64}), but for $m \neq 0$ it does not vanish identically
any more. In the presence of boundary conditions, $a^\pm (k)$ for positive, 
respectively negative energy in the S-matrix element (\ref{63}) are 
related to values $k \geq 0$. Each $k$ labels a mixture of left and 
right moving ``particles''. With (\ref{67}) the matrix element 
$T^{(S)}$ of (\ref{63}) is regular at ${\bar x}^0 = {\bar y}^0 = 0$. 
The same is true for the analogous one from $R^{(4)}$. In the 
latter the singularity ${\bar x}_0^{-2}$ is absent {\em only} for 
the solution a) ( $\alpha = 0$ in (\ref{42}) and (\ref{43}) ) thanks to the 
step functions. Both contributions to the total matrix element 
can be integrated completely yielding distributions. Some details 
of the calculation are described in Appendix A.

For ``incoming'' momenta $q, q'$ with energies $E_q, E_{q'}$ and 
outgoing ones $k, k'$ with $E_k, E^\prime_k$, taking into account 
symmetries and ensuing factors we obtain from (\ref{63})
\eqa{
T^{(S)} (q,q'; k,k') &=& \frac{m^8}{64} N(q)N(q')N(k)N(k') 
\de(E_k+E_k'-E_q-E_q') \nonumber \\
&& \left[ I(q,q'; k,k')+I(q,-k; -q',k')+\right. \nonumber \\
&& \left. +I(-k',q'; k,-q)+I(-k,-k'; -q,-q') \right]
}{68}
with
\eq{
I := \sum_{abrs} abrs K_{ab,rs}
}{69}
where the sum extends over $a, b, r, s,$ each taking the values 
$\pm 1$. With the abbreviations $( E = E_k + E^\prime_k = E_q + 
E^\prime_q$)
\eqa{
u_{ab} &:=& ak+bk'+E = \tilde{u}_{ab} + E \nonumber \\
v_{rs} &:=& rq+sq'+E = \tilde{v}_{rs} + E
}{70}
the quantity $K_{ab, rs} := \lim_{\eps \to 0} K (u_{ab}, v_{rs})$ is given by 
\eqa{
K(u,v) &:=& 2\pi i \frac{(u-v)^3}{u^2v^2} \ln{(u-v+i\eps)} \nonumber \\
&& + \frac{\pi i}{v^2}(3v-u) \ln{(u+i\eps)} \nonumber \\
&& - \frac{\pi i}{u^2}(3u-v) \ln{(-v-i\eps)} .
}{71}
The logarithms of complex arguments are defined as usual, i.e.:
\eq{
\lim_{\eps \to 0} \ln{(r \pm i\eps)} = \ln{\left|r\right|}\pm i\pi\Theta(-r)
}{logs}

The final result for the second contribution (\ref{61}) inserted in (\ref{63}) 
has a similar structure:
\eqa{
T^{(R)} &=& \frac{m^6}{16} N(q)N(q')N(k)N(k') \de(E_q+E_q'-E_k-E_k')
\nonumber \\
&& \left[ V(q,q'; k,k')+V(q,-k'; k,-q')+ \right. \nonumber \\
&& \left. +V(-k',q; k,-q)+V(-k,-k'; -q,-q')\right] \plabel{72} \\
V &:=& V^{(1)}+V^{(2)} \plabel{73} \\
V^{(1)} &:=& -\sum_{abrs} rs(E_q+rq)(E_q'+sq')J_{abrs} 
\plabel{74} \\
V^{(2)} &:=& \sum_{abrs} rsb(E_q+rq)(E_q'+sq')L_{abrs}
}{75}

In terms of $\tilde u_{ab}$ and $\tilde v_{rs}$ defined in (\ref{70}) 
we have for  $J_{abrs} := \lim_{\eps \to 0}J( \tilde u_{ab}, \tilde v_{rs} )$, 
$L_{abrs} := \lim_{\eps \to 0} L ( \tilde u_{ab}, \tilde v_{rs})$ the 
expressions 
\eqa{
J(\tilde{u},\tilde{v}) &:=& \frac{2\pi i}{\tilde{v}} \ln{(\tilde{u}+\tilde{v}
+i\eps)} \plabel{76} \\
L(\tilde{u},\tilde{v}) &:=& -2\pi i\frac{(E-\tilde{v})}{(E-\tilde{u})^2}
\ln{(E-\tilde{v}-i\eps)}
}{77}

For both contributions the common overall sign has been fixed by 
the choice $2 \sqrt{2} {\bar q}_3 = +1$ in (52). Both terms also 
share the energy conservation factor. Therefore, only a 
probability per unit of time with factor $ \frac{1}{2 \pi}\, 
\delta (E_k + E'_k - E_q - E'_q )$ is a well defined quantity. 
The situation with respect to momenta is different because of the nonlocality
of the vertex. Thus we encounter a situation similar to scattering at a fixed
external ``potential'' in ordinary quantum mechanics (or, equivalently, in 
$D=0+1$ dimensional QFT). 

It should be noted that in the infrared limit both amplitudes are proportional 
to $m$ and hence vanish for the massless case in agreement with the previous 
discussion. For large energies\footnote{Note that for very large 
energies our perturbation theory breaks down since the effects from the scalar 
field are not ``small'' anymore; therefore, the energy should lie in the range
$m \ll E \ll E_{Planck}$.} the amplitudes decrease rapidly:
\eq{
\left. T^{(S)} \right|_{UV} \propto m^8 \frac{\ln E}{E^7}, \hspace{1cm} 
\left. T^{(R)} \right|_{UV} \propto m^6 \frac{\ln E}{E^5}
}{n2}

\section{Summary and Outlook}

Two dimensional path integral quantum gravity can now be based upon a 
well-defined formalism which --- in a very specific gauge --- 
allows to separate the exact, almost trivial, quantum integral of 
the geometric variables from the loop-wise effects of the 
scalars. In our present work we considered the classical, 
tree-approximation, limit for minimally interacting scalar fields $S$, 
starting from the path integral formalism. This implies the appearance of 
effective (classical) $2n$-vertices of scalar fields ($n \geq 
2$). Those vertices are highly nontrivial, because they yield --- 
through the natural appearance of classical background phenomena 
--- mathematical structures which allow the interpretation that 
an intermediate ``virtual'' BH is involved. We have 
studied this for the geometric action as derived by spherically 
reducing Einstein gravity. The scalar matter field was assumed to 
be coupled minimally at the $d = 2$ level. We also concentrated 
upon the simplest nontrivial vertex $S^{(4)}$ with four scalar 
fields.

For the massless case we found that the resulting nonlocal matrix 
element for unrestricted left- and right-moving scalar fields 
diverges as $\int_0^{z_f} \frac{dz}{z^2}$ at that point in space-time which 
can be identified with the ``location'' of a singularity. However, 
imposing a suitable boundary condition upon $S$ completely 
eliminates the scalar excitation moving towards the singularity, 
whereas the ones moving away decouple from the theory: The 
manifold has been ``plugged'' at the place where an eventual 
BH may have been formed. We believe that this result at 
the particle level shows a qualitative relation to a conjecture
for macroscopic BH formation: There for minimally coupled scalars a 
BH forms without any threshold \cite{chop}. The divergence of a 
probability amplitude or, 
alternatively, an amplitude which is finite only for a wave 
packet, properly rescaled to tend to zero width at $z =0$, seems to 
imply the same phenomenon. 

As an example for a system where finite amplitudes for minimally coupled 
scalars can be obtained we also 
studied massive scalars, where the necessary boundary 
condition no longer prevents BH effects. Both, the vertex 
from the massless case, and another new one, induced by the 
mass-term, yield finite results which can be even represented by 
(complicated) sums of directions (plus or minus) of momenta in 
terms of (simple) functions and distributions. Overall energy 
conservation holds in the process $S + S \to S + S$. Momenta are 
not conserved, in general. Here we note parallels to recent work 
of P. H\'aj\'i\v cek \cite{haj99} on massless, but non-minimally coupled thin
spherical shells. He found no residual BH 
for a collapsing shell with Dirac quantization. Also in that work the 
phenomenon has been observed which we have called ``virtual BH'', consisting 
in a certain sense of a black and a white hole.

The next task \cite{gkv99} is to take into account also the proper 
nonminimal coupling of scalars at the $2d$ level. Superficially 
no essential basic changes for the vertices may be expected: On 
the one hand, the measure of the integral will change as $dz d\tau 
\to z^2 dz d\tau $, because $z$ will become a radial variable. On 
the other hand the scalar field will be reduced to the one 
describing $s$-waves in $d = 4$, i.e.\ $S \to S/z$. But e.g.\ the 
threshold effect known for macroscopic studies \cite{cho93} should show 
up. In that case a detailed comparison with Dirac quantization as treated in
\cite{haj99} will be possible.

Our formalism is general enough so that any other 2d gravity 
theory, produced e.g.\ by spherical reduction of generalized 
Einstein gravities in $d =4$, can be covered as well.

Of course, also the study of higher loop orders in the scalar 
fields, based upon the one-loop determinant (Polyakov type 
action) in the path integral, as well as of higher loops involving 
the vertices discussed here, together with propagators of the 
scalars, remains a wide field of possible further applications.

\section*{Acknowledgment}

The authors thank their colleagues at the Institute for 
Theoretical Physics of the Vienna University of Technology for 
many discussions. This paper has been supported by project P-12815-TPH of the 
Austrian Science Foundation (FWF). One of the authors (D.V.) has been 
supported in part by the Alexander von Humboldt foundation and by RFBR, grant 
97-01-01186.

\section*{Appendix A: Derivation of Scattering Amplitudes}

The explicit computation of the scattering amplitude $T^{(S)}$ in 
(\ref{68}) and $T^{(R)}$ in (\ref{72}) for Minkowski modes in the initial and 
final state is most conveniently based upon suitable Fourier transforms of the 
rational factors in ${\bar x}^0$ and ${\bar y}^0$ in (\ref{60}) and (\ref{61}).
From the identity \cite{gelfand} 
\eq{
\int_0^{\infty} x^{\la} e^{i(\si+i\eps)} dx = i e^{\frac{i\la\pi}{2}} \Gamma
(\la+1) (\si+i\eps)^{-\la-1}
}{a1}
the required singular limits $\delta \to +0, \eps \to +0$ for the Fourier 
transforms at $\lambda = -2 + \delta$, resp.\ $\lambda = -1 + \delta $ are
\eq{
\int_{-\infty}^{\infty} x^{-2+\de}\Theta(x)e^{i(\si+i\eps)} = f_2(\si,\eps) 
\left[\left(\frac{1}{\de}+(1+\frac{i\pi}{2}-\ga)\right) - \ln{(\si + i\eps)}
+ {\cal O}(\de)\right]
}{a2}
resp. 
\eq{
\int_{-\infty}^{\infty} x^{-1+\de}\Theta(x)e^{i(\si+i\eps)} = f_1(\si,\eps) 
\left[\left(\frac{1}{\de}+(\frac{i\pi}{2}-\ga)\right) - \ln{(\si + i\eps)}
+ {\cal O}(\de)\right] ,
}{a3}
with
\eq{
f_n := i(-1)^{(n-1)}e^{-i\frac{n\pi}{2}}\left(\si+i\eps\right)^{(n-1)}
}{anew} 
\begin{figure}
\centering
\epsfig{file=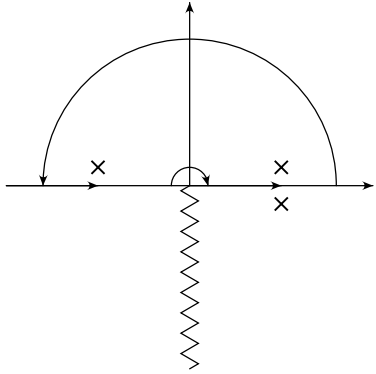, height=5cm}
\caption{The complex contour typically encircles 2 singularities; 
note that for convenience we put the branch cut for the logarithm on the 
negative imaginary axis (depicted by the zigzag line).}
\plabel{fig3}
\end{figure}
Introducing (\ref{a2}), resp.\ (\ref{a3}) into (\ref{68}) resp.\ (\ref{72}) 
and using ($P$ is Cauchy's principal value)
\eqa{
\eps(x) = \frac{1}{\pi i}P\int d\tau \frac{e^{i\tau x}}{\tau} \nonumber \\
\Theta(x) = \frac{1}{2\pi i}\int d\tau \frac{e^{i\tau x}}{\tau-i\eps}
}{a4}
all the terms proportional to $\delta^{-1}$ from (\ref{a2}) and (\ref{a3}) 
cancel together with the constant contributions\footnote{These cancellations 
are a direct consequence of the boundary condition (\ref{64}).} --- as 
they should in these finite integrals. Furthermore in (\ref{68}) it is 
useful to replace the factor $({\bar x}^0 - {\bar y}^0)^3$ by a 
third derivative with respect to $E = E_k + E^\prime_k$. Then the 
generic integral 
\eq{
A (a,b) = \int d\tau \ln(a+\tau+i\eps_1) \ln(-b-\tau+i\eps_2) (a+\tau) (b+\tau)
(\frac{1}{\tau+i\eps_3}+\frac{1}{\tau-i\eps_3})
}{a5}
remains which after three differentiations with respect to $E$ 
in $a = E + q + q', \; b = E + k + k'$ becomes a 
contribution of integrals with one logarithm multiplied by a factor 
with two or three poles. These integrals are straightforward and can be
most conveniently done using the contour depicted in Fig. \ref{fig3}. 
\\
In the vertex $T^{(R)}$ in (\ref{72}) for the first contribution 
$V^{(1)}$ the procedure is the same, not even requiring some 
differentiation at an intermediate step. $V^{(2)}$ originates 
from the term with factor $(-{\bar y}^0 {\bar x}^{-2}$). Here 
${\bar y}^0$ may be expressed first by a derivative with respect to one 
of the momenta in the sine factor from $(\partial_{\bar y} S)^2$ 
(cf.\ (\ref{67})).

\end{document}